# Terahertz emission and detection using Ge-on-Si photoconductive antennas

Dhanashree Chemate[1,2,3], * Abhishek Singh[4], * Utkarsh Pandey[2], Ruturaj Puranik[2], Vivek Dwij[2], Priyanshu Gahlaut[4], Siddhartha P. Duttagupta[5], and Shriganesh S. Prabhu[2]

[1]Department of Metallurgical Engineering and Materials Science, Indian Institute of Technology Bombay, Mumbai, 400076, India

[2]Department of Condensed Matter Physics and Materials Science, Tata Institute of Fundamental Research, Mumbai, 400005, India

[3]Centre for Interdisciplinary Sciences, Tata Institute of Fundamental Research, Hyderabad, 500107, India

[4]Centre for Advanced Electronics, Indian Institute of Technology Indore, Indore, 453552, India

[5]Department of Electrical Engineering, Indian Institute of Technology Bombay, Mumbai, 400076, India

Email: dhanashree.c@iitb.ac.in, asingh@iiti.ac.in

## ABSTRACT

Germanium-on-Silicon (Ge-on-Si) is a promising, CMOS-compatible platform for integrated terahertz (THz) photonics, offering a low-cost alternative to III-V semiconductors. A primary challenge for Ge-based photoconductive antennas (PCAs), however, has been the long carrier lifetime of bulk Ge, preventing its use as a detector. Here, we demonstrate that amorphous Ge (a-Ge) films overcome this limitation, possessing inherent ultrashort carrier lifetimes ~ 1.11-1.38 ps. We leverage this property to demonstrate, for the first time to our knowledge, coherent THz pulse detection using undoped a-Ge-on-Si PCAs. We present a comparative study of devices fabricated on a-Ge films grown by plasma-enhanced chemical vapor deposition (PECVD) and DC magnetron sputtering. The PECVD-Ge device, with better homogeneity and a smoother morphology in the films, demonstrates superior performance for both THz emission and detection. As an emitter, the PECVD-Ge PCA achieves a 40 dB signal-to-noise ratio (SNR) with a bandwidth of ~ 3 THz. As a detector, it achieves a 32 dB SNR and a ~ 2 THz bandwidth, representing a ~2.5-fold increase in detected signal amplitude over the sputtered-Ge device. These results establish amorphous Ge-on-Si as a viable and scalable platform for both THz generation and detection, paving the way for fully integrated Si-based THz time-domain systems.

## 1. Introduction

The terahertz (THz) frequency range (0.1-10 THz) of electromagnetic radiation offers opportunities for diverse applications, including material studies, medical imaging, security screening, non-destructive testing, and high-speed communication [1-4]. However, the widespread adoption of THz technology is currently limited by the need for compact, efficient, and cost-effective THz sources and detectors [5]. Thus, the research to develop such systems is gaining significant momentum [6]. Photoconductive antennas (PCAs) are a popular method for generating and detecting terahertz (THz) radiation, allowing control over the output power and polarization through the applied bias and electrode design [7-10]. Furthermore, their compact size makes PCAs a promising candidate for the integrated THz systems with a small footprint.

Among the widely studied photoconductive materials for THz sources and detectors are the III-V-based compound semiconductors, mainly GaAs and InGaAs, which offer excellent optoelectronic properties [11]. The photoconductive devices require femtosecond (fs) NIR pulses to photoexcite the electrons from the valence band to the conduction band in the semiconductor. Typically, GaAs-based THz devices are operated with 800 nm fs lasers, and InGaAs-based devices are operated with 1550 nm fs lasers. This led to the formation of mainly two different material-laser platforms, GaAs-Ti:Sa:800 nm and InGaAs-Er fiber:1550 nm, for the development of photoconductive THz technology. With extensive research spanning more than three decades on these two platforms, significant advancements in photoconductive THz technology have been achieved [12-17]. The THz electric field ~ 230 kV/cm [18], average THz powers in several mW [19], gapless spectral bandwidth up to 7 THz in regular transmission mode [20], up to 20 THz with absorption dips in the reflection mode of the emitter [21], and a dynamic range of 137 dB has been reported [22]. However, there are several challenges involved in the fabrication of these materials due to the complex growth techniques, such as molecular beam epitaxy (MBE) [23,24]. Additionally, with the presence of a polar phonon mode around 8 THz, the spectrum exhibits significant absorption dips at these frequencies [25-27].

On the other hand, with the onset of integrated mode-locked lasers, fibre-optic lasers with readily available telecom wavelengths for on-chip ultrafast optical systems [28,29] and the demand for cost-effective, compact THz sources, propels us to study the other group of materials for PCAs. In recent years, Germanium (Ge) has emerged as a promising PCA material, mainly because of the absence of polar phonons and its compatibility with both CMOS integration and a wide range of lasers with a wavelength range < 800 nm to 1500 nm [23-27]. It was shown that despite a longer carrier lifetime (~2 ns) in bulk Ge implanted with gold (Au), the THz emission bandwidth reached up to 70 THz [26]. Along with the absence of polar phonons in Ge, the key factor contributing to this promising result was the short pulse width of the excitation laser, which was ~ 11 fs. Other studies on Ge-PCA involve Ge implanted with phosphorus, along with the silicon waveguide, which showed emission up to 1.5 THz [24]. Among the thin-film-based devices, the study on thermally evaporated GeSn achieved a THz emission bandwidth of up to 1.5 THz with a signal-to-noise ratio (SNR) of 40 dB, utilizing plasmonic gratings embedded in the bowtie [23]. A large-area PCA on these films showed up to 2 THz emission with a 40 dB SNR [27]. Other than PCAs, Ge-on-Si devices are also gaining attention in biomolecule sensing applications using plasmonics [30, 31]. In these studies, the sensing of a deep subwavelength layer of an antioxidant and protein is demonstrated using an array of Ge bowtie antennas with resonance frequencies between 500 GHz and 1 THz. Moreover, Ge has been chosen over the other III-V group semiconductors due to its CMOS compatibility, despite the latter having inherently higher mobilities. The use of Ge allows for compact sensors with mass production capacity, which are essential for consumer electronics.

However, even with these developments, Ge remains one of the least explored materials in THz photonics applications, especially among the photoconductive materials for THz PCAs. While Ge-based emitters have shown promise, to our knowledge, THz pulse detection has not yet been reported with undoped Ge. Doped Ge has been used to develop THz photodetectors [32], as an incoherent detector and, more recently, in coherent detection as well [33]. The long (a few μs) carrier lifetime of bulk Ge has remained a fundamental roadblock for its use in coherent THz detection, which requires sub-picosecond carrier dynamics [34]. To overcome this, most work has focused on ion implantation or alloying with Sn [23,27,33]. However, amorphous Ge (a-Ge) inherently possesses a high density of trapping sites, which naturally quenches the carrier lifetime to the sub-picosecond regime [35,36]. This property, often considered a detriment to electronic performance, is precisely one of the requirements for an effective PCA material.

In this study, we investigate the potential of a-Ge thin films, grown by two scalable, CMOS-compatible methods-plasma-enhanced chemical vapor deposition (PECVD) and DC magnetron sputtering-for THz PCA applications. We demonstrate the first coherent THz pulse detection using a-Ge-on-Si PCAs, enabled by the material's intrinsic ultrashort carrier lifetime. We then provide a comprehensive comparison of the material properties (structural, optical, and electrical) of the films from both deposition methods and correlate them directly to device performance. We show that the PECVD-Ge device, possessing better film homogeneity, yields significantly higher efficiency for both THz emission and detection, establishing a clear path toward optimized, low-cost, and integrated THz systems on a silicon platform. The devices are addressed as Ge-on-Si PCA, and for the identification with deposition methods, we address them as PECVD-Ge device or sputtered-Ge device.

## 2. Material characterization and device fabrication

The Ge thin films grown by PECVD and magnetron sputtering are characterized using Grazing Incidence X-ray diffraction (GI-XRD), Raman spectroscopy, resistivity measurements, optical pump terahertz probe spectroscopy (OPTP), energy dispersive X-ray spectroscopy (EDAX), scanning electron microscopy (SEM), and atomic force microscopy (AFM). The PECVD-Ge is commercially obtained from CENSE, IISc, Bangalore. With a thickness of ~ 1 µm, the film is deposited on a high-resistive silicon (HR-Si) wafer ($\rho$ ~ 10000 $\Omega$-cm). Sputtered - Ge film of similar thickness is deposited using DC magnetron sputtering with a power of 40 W and a sputtering pressure of 8 x $10^{-3}$ mbar. Prior to deposition, a base vacuum of 8 x $10^{-6}$ mbar is achieved. A consistent deposition rate of 0.44 nm/s is maintained throughout the deposition process, and the deposition is carried out without any substrate heating. The GI-XRD measurements, as presented in Figure S1 in the supplementary information, show the presence of broad peaks around 20°-30° and 40°-60°, which predominantly show an amorphous nature of the deposited films. To further confirm this, Raman Spectroscopic measurements were carried out using a WITec alpha 300R confocal microscope with a 532 nm excitation wavelength and 500 µW laser power. Figure 1 shows a sharp peak at 301 $cm^{-1}$ for c-Ge, whereas both PECVD and sputtered-Ge films show broader peaks at 275 $cm^{-1}$, indicating disordered, amorphous Ge [37]. It should be noted that a small shoulder peak in the PECVD-Ge indicates that the films could consist of a nanocrystalline formation.

Further, the roughness of the films is measured using AFM, where the PECVD-Ge film with a roughness of ~ 0.48 nm [Figure. S2(a)] is smoother than sputtered-Ge films with a roughness of ~ 4.9 nm [Figure. S2(b)]. This distinct difference in the morphology of the films is consistent with the energetics of the two different deposition techniques. In sputter deposition, atoms ejected from the target typically possess more kinetic energies than the gas species in PECVD, leading to relatively higher roughness of the films [38].

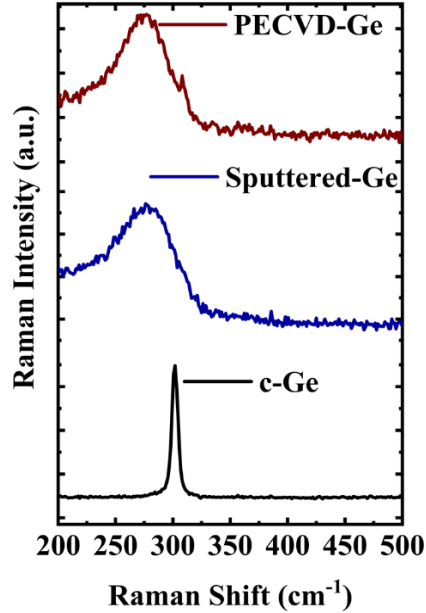

**Figure 1.** Raman spectra of PECVD-grown Ge thin film, sputtered Ge thin film, and crystalline (c-Ge). The broad peaks in the PECVD and sputtered Ge are shifted to 275 cm$^{-1}$, confirming the amorphous nature of the films

Next, to study the carrier dynamics of the Ge films, the OPTP measurements are performed using an Amplifier laser ($\lambda$ = 800 nm, repetition rate: 1 kHz, power: 5 W, pulse width ~ 35 fs). A pump power of 200 mW (intensity 2 mJ/cm$^2$) is used for the excitation of the samples. The relative change in THz transmission is measured and is shown in Figures 2(a) and 2(b), indicating an exponential decay in both PECVD-Ge and sputtered-Ge samples, respectively. Equation (1) is used to fit the experimentally measured data points of exponential decay, where $t_o$ corresponds to time zero of the exponential fit, and $\tau_{decay}$ is the decay time constant. $A_0$ is a constant, and $A_1$ is the amplitude of the decay function [35,36]. The decay constants of PECVD-Ge and sputtered-Ge samples are calculated to be around 1.38 ± 0.01 ps and 1.11 ± 0.02 ps, respectively. A relatively faster recombination in sputtered-Ge samples is potentially due to differences in defect density or trapping sites within the films.

$$-\frac{\Delta T}{T}(t) = A_o + A_1 e^{-\frac{t-t_o}{\tau_{decay}}} \tag{1}$$

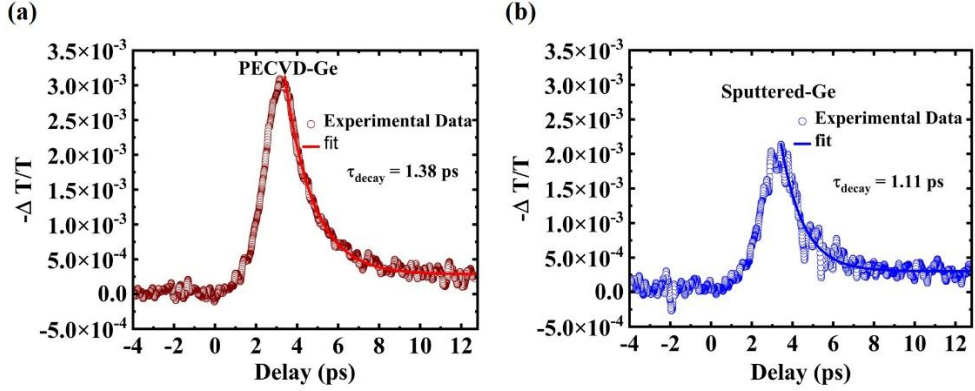

**Figure 2**. Carrier lifetime in (a) CVD-Ge film ($\tau_{decay}$ = 1.38 ps) and (b) Sputtered-Ge ($\tau_{decay}$ = 1.11 ps) film measured using OPTP. The carrier lifetime of the PECVD-Ge film is slightly longer, potentially due to the superior quality of the films.

To further study the electrical properties, resistivity measurements are carried out using the van der Pauw method, with the obtained values presented in Table 1. The PECVD-Ge films ($\rho$ ~8.58 ± 0.06 $\Omega$.cm) exhibit a significantly lower resistivity (higher conductivity) than the sputtered-Ge films ($\rho$ ~ 60.03 ± 1.54 $\Omega$.cm). This is consistent with the other material characterizations, where the PECVD-Ge films are more homogeneous (as shown in the SEM images in Figure S3), smoother (Figure S2), and contain fewer impurities, as seen in the EDAX data in Figure S4. The longer carrier lifetime in PECVD-Ge (Figure 2) also suggests a relatively lower density of deep-level defects compared to the sputtered film.

**Table 1: Resistivity measurements of Ge thin films of ~ 1 µm thickness**

| Sample | Sheet resistance ($\Omega$/sq) | Resistivity ($\Omega$.cm) |
|---|---|---|
| PECVD-Ge | $(8.58 \pm 0.01) \times 10^4$ | $8.58 \pm 0.01$ |
| Sputtered-Ge | $(6.00 \pm 0.09) \times 10^5$ | $60.0 \pm 0.9$ |

Following the characterization of the films, two identical PCAs are fabricated on both samples using a standard photolithography technique and a lift-off method. Figure 3 shows the optical microscopic image of the device along with the scanning electron microscopic image of the fabricated PCA. The antenna electrodes consist of AuGe metal alloy of ~ 150 nm thickness, and the gap between the electrodes is maintained at around ~21.4 µm.

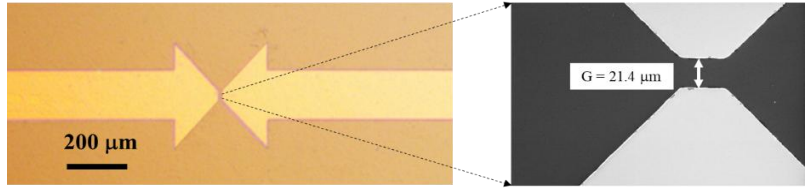

Figure 3. Optical microscopic image (left) of the fabricated PCA on Ge thin films with the magnified SEM image (right) showing the gap between the electrodes to be around 21.4 μm

## 3. Experiment

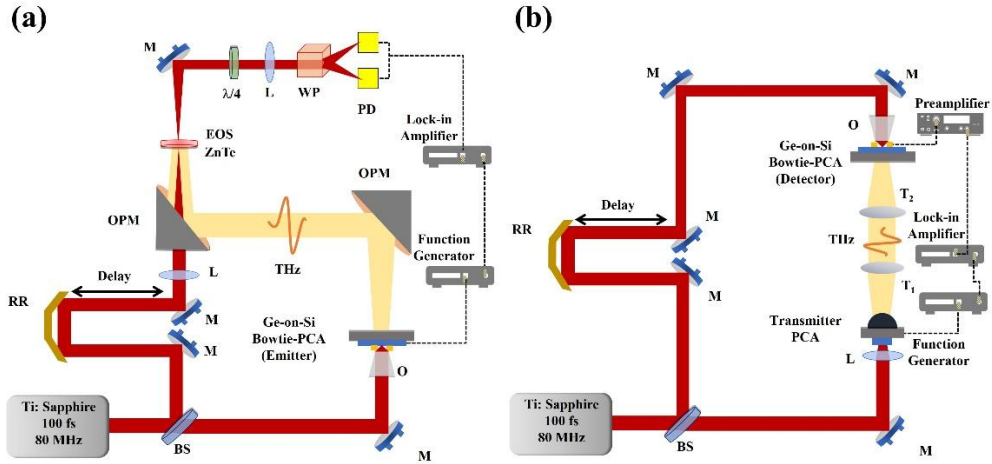

**Figure 4.** (a) Schematic of the experimental setups for the characterization of Ge-on-Si PCAs (a) THz generation, (b) THz detection. M: Mirror, RR: Retroreflector, L: Lens, PD: Balanced Photodiode, OPM: Off-axis Parabolic Mirror, WP: Wollaston Prism, BS: Beam Splitter, O: Objective, T: TPX lens

In order to study the devices as THz emitters, we used a standard THz time-domain spectroscopy (TDS) setup as illustrated in Figure 4(a). The laser used is a mode-locked femtosecond Ti: Sapphire with a repetition rate of 80 MHz and a pulse duration of 100 fs. The laser beam is split into a pump and probe path, where the probe beam is delayed. A 10x objective is used to focus the pump beam in the gap of the bowtie antenna. The device is biased using a function generator with a chopping frequency of 27.437 kHz. Two parabolic mirrors are used to collimate and focus the THz beam. At the detection, an electro-optic sampling is employed with the 1 mm thick ZnTe <110>. The signal from the balanced photodiodes is sent to the lock-in amplifier to measure the emission response of the Ge-on-Si PCA.

The same devices are characterized as THz detectors in the setup shown in Figure 4(b), for which we employed a Batop GmbH iPCA-21-05-1000-800-x-PCA as the source. This PCA is biased with a 10 V peak-to-peak voltage along with a chopping frequency of 27.437 kHz. The probe beam is focused and incident on the device in the gap of antennas, whereas THz is illuminated from the back of the antenna. The response from the detector Ge-on-Si PCA is further fed into a preamplifier and a lock-in amplifier.

## 4. Results and discussion

The current-voltage (I-V) measurements are carried out in dark and illuminated conditions with varying powers of 10, 20, and 30 mW of 800 nm, 100 fs, 80 MHz pump. The dark characteristics in Figure 5(a) show that, albeit with an identical design, PECVD-Ge devices exhibit a dark resistance an order lower (19.8 kΩ) than that of sputtered-Ge devices (250 kΩ). The larger dark current flowing through PECVD-Ge devices is consistent with the difference in material properties observed in the samples with resistivity measurements. On the contrary, the photocurrents are higher in sputtered-Ge (Figure 5(c)) than PECVD-Ge (Figure 5(b)), potentially because of the higher effective absorption of the incident light due to the increased roughness [39]. The inhomogeneity of the sputtered film may also have contributed to the increased Si substrate photo response, as more light was able to penetrate and reach the Si substrate [40].

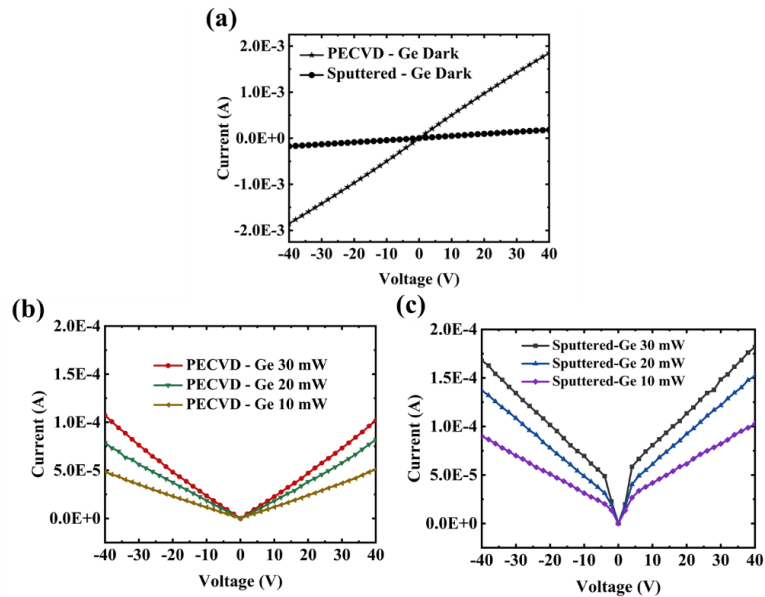

**Figure 5.** The current-voltage characteristics (a) Dark characteristics of the PCAs on PECVD-Ge and sputtered-Ge films (b) Photocurrent (total current-dark current) of a PCA on PECVD-Ge (c) on sputtered Ge. For improved clarity and direct comparison, the photocurrents are plotted on the positive y-axis.

To study the performance of the Ge-on-Si PCAs as THz sources in the TDS setup, the PCAs are biased with 40 V peak-to-peak voltage and illuminated with 800 nm, 100 fs, 80 MHz pump and 30 mW power with a spot size of ~ 30 µm. With a fluence of ~ 106 µJ/cm², the device is operated well below 26 mJ/cm² - the femtosecond laser-induced crystallization limit for amorphous Ge films [41]. The PCA as a source works on the principle that when the gap of the antenna is illuminated by the optical pump beam, the carriers are generated in the photoconductive material. These carriers are accelerated in the presence of a device bias, which further gives rise to the electric field that is given as $E_{THz} \propto \frac{dJ}{dt}$, where J is the current flowing through the device. Figure 6(a) shows the THz time domain pulse generated by PCAs on PECVD-Ge and sputtered-Ge films. The peak signal for the PECVD-Ge device is 600 µV, whereas it is 450 µV for the sputtered-Ge device. The corresponding FFT in Figure 6(b) shows the THz bandwidth reaching up to 3 THz with an SNR of 40 dB and 36 dB for PECVD-Ge and sputtered-Ge PCAs, respectively. The PCA on PECVD-Ge shows better performance, potentially due to the relatively higher mobility, which is a result of the homogenous film. Also, the parasitic current contribution from the Si substrate possibly leads to slightly lower THz peak values in the sputtered-Ge device than in the PECVD-Ge devices [40].

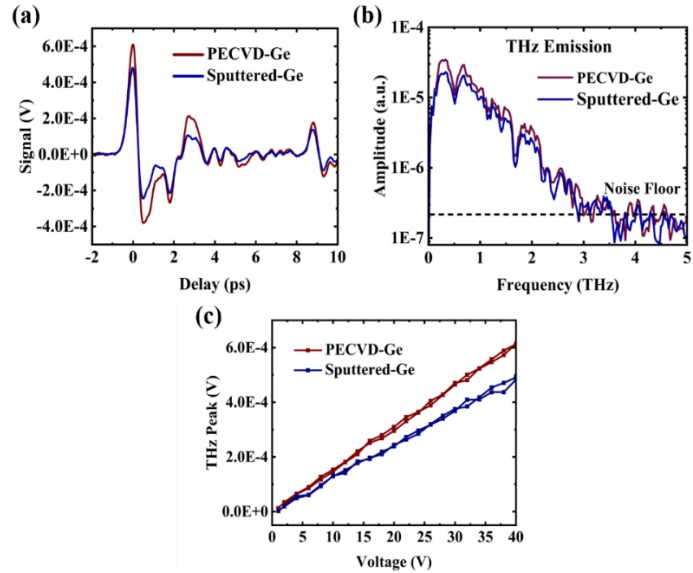

**Figure 6.** THz generation from the bowtie PCAs on PECVD-Ge and sputtered-Ge films (a) Time-domain and (b) frequency domain signal of the generated terahertz pulse for 30 mW at 40 V bias. The PECVD-Ge device shows slightly higher SNR.

(c) THz peak vs applied bias with forward and backward sweep. The devices show stable performances, with no significant change in the THz peak values.

Further, to check the stability of the devices, a forward and reverse sweep up to 40 V is given, and the THz peak is recorded; it can be seen in Figure 6(c) that the signal is constant in both the devices on PECVD-Ge and sputtered-Ge. Thus, showing the potential of Ge thin films for a stable output in THz generation. The performance metrics of Ge-PCAs as emitters are summarized in Table 2.

Table 2: Comparative THz-PCA Performance Metrics: Emitter

| PCA Material | Peak Signal (µV) | SNR (dB) | Bandwidth (THz) |
|---|---|---|---|
| PECVD-Ge | 600 | 40 | 3 |
| Sputtered-Ge | 450 | 36 | 3 |

For the characterization of devices as detectors, the THz beam, generated from a commercial PCA, is focused from the back of the substrate, and the probe beam with a power of 60 mW is focused in the gap of the bowtie antenna. The carriers are generated by the probe beam, which are accelerated due to the field induced by the THz pulse. The current generated in the device is directly proportional to the strength of the field [42, 43]. This signal is further amplified with a gain of 200 nA/V. Figure 7(a) shows the THz pulse detected with the Ge PCAs, which is emitted from a commercial PCA. The peak of the PECVD-Ge device is significantly higher (~100 µV) than that of the sputtered-Ge PCA (~40 µV). Similarly, the FFT in Figure 7(b) shows the THz spectrum up to ~2 THz with a 32 dB and 24 dB SNR for PECVD-Ge and sputtered-Ge PCAs, respectively. These detector performance metrics are summarized in Table 3. To the best of our knowledge, this is the first reported broadband THz detection bandwidths from undoped a-Ge-on-Si PCAs. It should be noted that the devices are characterized in ambient air and without the integration of a Si lens with the substrate. The reflection losses from the Si-air interface could contribute significantly to such sensitive measurements. Thus, it is expected that the SNR can be further improved with the ideal experimental conditions. Additionally, a shorter pulse width results in a faster rise of the photo-induced charge carriers, thereby dominating the transient current [26]. Thus, a shorter pulse duration can be utilized to exploit the full potential of Ge-on-Si PCAs as both THz emitters and detectors in terms of bandwidth. Overall, these results show the promise of Ge-on-Si technology for the emission and detection of THz pulses.

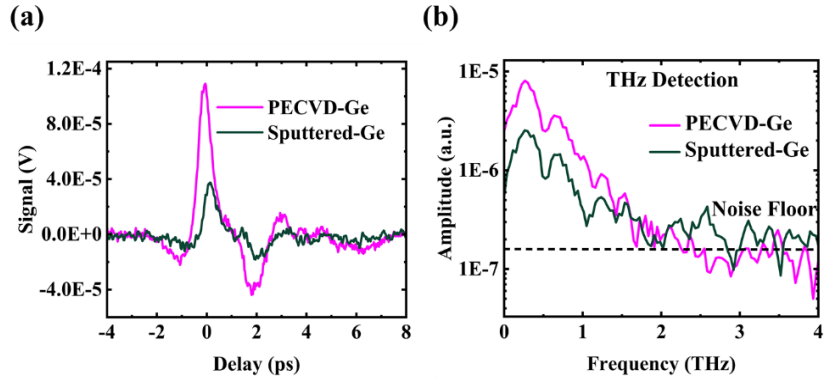

**Figure. 7.** THz detection from the bowtie PCAs on PECVD-Ge and sputtered-Ge films (a) Time-domain and (b) frequency domain signal of the detected terahertz pulse. The PECVD-Ge PCA shows higher SNR.

**Table 3: Comparative THz-PCA Performance Metrics: Detector**

| PCA Material | Peak Signal (μV) | SNR (dB) | Bandwidth (THz) |
|---|---|---|---|
| PECVD-Ge | 100 | 32 | 2 |
| Sputtered-Ge | 40 | 24 | 2 |

## 5. Conclusion

In conclusion, we have demonstrated the viability of amorphous Ge-on-Si thin films as a complete, CMOS-compatible platform for THz photoconductive antennas. We have shown that amorphous Ge, grown by both PECVD and sputtering, possesses the key enabling property for THz detection: an intrinsic ultrashort carrier lifetime (1.11-1.38 ps). Leveraging this, we have presented the first demonstration of coherent THz pulse detection using a-Ge-on-Si PCAs. A comparative study found that the deposition method is critical to device performance. PECVD-grown films exhibit superior structural homogeneity and lower surface roughness. This translated directly to superior device metrics: the PECVD-Ge PCA achieved a 40 dB SNR and ~ 3 THz bandwidth as an emitter, and a 32 dB SNR and ~ 2 THz bandwidth as a detector. This work addresses a key bottleneck in Si-based THz systems by providing a low-cost, scalable material that can function as both an emitter and a detector. While further optimization of material growth and device design is possible, this study paves the way for the monolithic integration of full THz time-domain spectroscopy systems on a single silicon chip.

**Funding.**

Department of Atomic Energy, India, vide grant RTI4003 and SERB India under project RJF/2022/000081.

**Disclosures.**

The authors declare no conflicts of interest.

**Data availability.**

Data underlying the results presented in this paper are not publicly available at this time but may be obtained from the authors upon reasonable request.

**Supplementary Information.**

See Supplementary Information for supporting content.

Supplementary Information

**Terahertz emission and detection using Ge-on-Si photoconductive antennas**

Dhanashree Chemate[1,2,3], * Abhishek Singh[4], * Utkarsh Pandey[2], Ruturaj Puranik[2], Vivek Dwij[2], Priyanshu Gahlaut[3], Siddhartha P. Duttagupta[5], and Shriganesh S. Prabhu[2]

[1]Department of Metallurgical Engineering and Materials Science, Indian Institute of Technology Bombay, Mumbai, 400076, India

[2]Department of Condensed Matter Physics and Materials Science, Tata Institute of Fundamental Research, Mumbai, 400005, India

[3]Centre for Interdisciplinary Sciences, Tata Institute of Fundamental Research, Hyderabad, 500107, India

[4]Centre for Advanced Electronics, Indian Institute of Technology Indore, Indore, 453552, India

[5]Department of Electrical Engineering, Indian Institute of Technology Bombay, Mumbai, 400076, India

**Material Characterization**

In order to study the nature of the Ge thin films, the Grazing incidence X-ray diffraction data (GI-XRD) measurements were carried out. Figure S1 shows the data recorded with the 2θ angle varied from 10° to 80° while maintaining a fixed incidence angle (ω) of 1.00°. The broad peaks around 20°- 30° and 40°- 60° show the amorphous nature of the films.

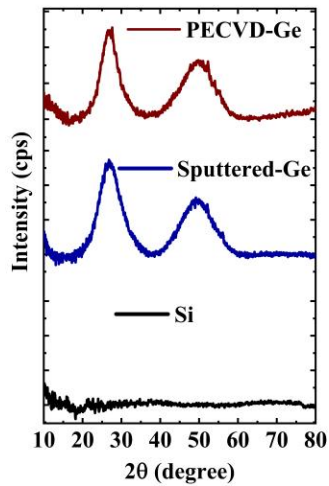

**Figure S1.** The GI-XRD of PECVD-Ge and Sputtered Ge along with the substrate Si. The broad peaks around 20º-30º and 40º-60º confirm the amorphous nature of the Ge films

Figure S2(a) and Figure S2(b) show the Atomic Force Microscopy (AFM) scans of PECVD-Ge and sputtered-Ge films, respectively. The roughness in the sputtered film is observed to be more than that of PECVD-Ge films, as typically expected for a more energetic deposition technique.

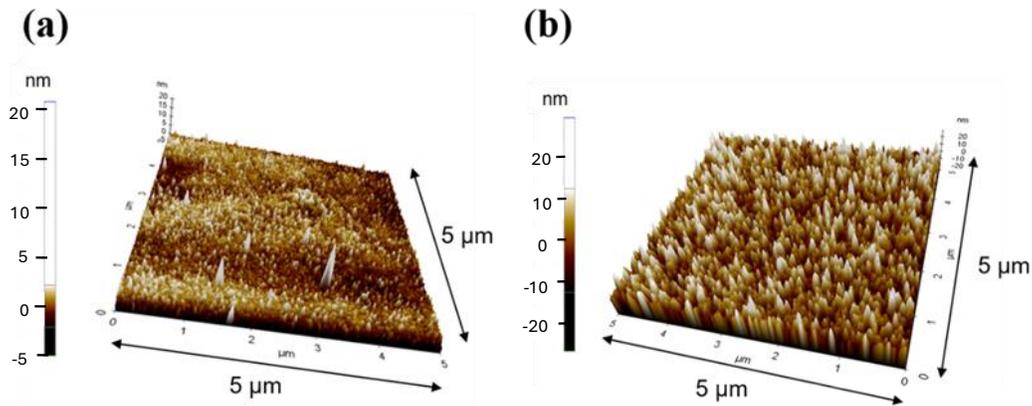

**Figure S2.** The AFM scan of (a) PECVD-Ge (b) Sputtered-Ge; the average roughness Ra of PECVD is 0.48 nm, whereas the sputtered film has an Ra of 4.8 nm

Figure S3 (a) and Figure S3(b) show the scanning electron microscopic images (SEM) of PECVD-Ge and sputtered-Ge films, respectively. The PECVD-Ge films showed more homogeneity, while the sputtered films looked relatively porous, which could be related to the presence of other constituents such as oxygen in the films. This difference is also reflected in the resistivity measurements of the films, with PECVD-Ge films showing higher conductivity than the sputtered-Ge.

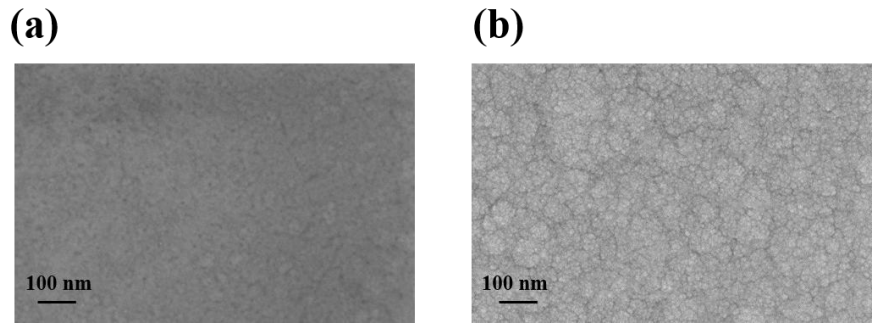

**Figure S3.** SEM images of (a) PECVD-Ge (b) Sputtered-Ge films; the PECVD-Ge films are more homogenous with the average grain size of 20.24 nm, whereas the sputtered-Ge film has a smaller grain size of 16.7 nm and a relatively porous nature

Figure S4(a) and Figure S4(b) show energy dispersive X-ray spectroscopy (EDAX) data of PECVD-Ge and sputtered-Ge films, respectively. The data show that Ge content in the PECVD-

Ge deposited film is relatively higher than that of the sputtered film. We attribute this difference to the different base vacuum levels in both techniques. Although quantifying oxygen using EDAX is challenging, the spectrum provides a comparative picture of the percentage of Ge constituents in both films.

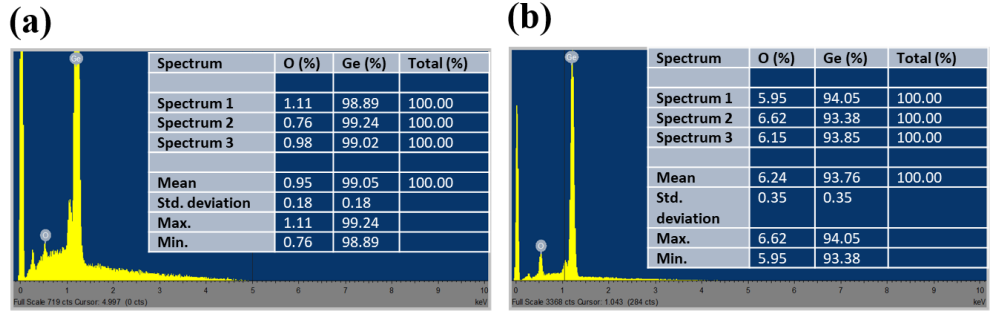

**Figure S4.** EDAX spectra of (a) PECVD-Ge films, (b) Sputtered-Ge films. The oxygen content is relatively higher in the sputtered films, pertaining to the base vacuum levels of the deposition chamber